\documentclass[fleqn,usenatbib,letters]{mnras}

\usepackage{newtxtext,newtxmath}
\usepackage[T1]{fontenc}

\DeclareRobustCommand{\VAN}[3]{#2}
\let\VANthebibliography\thebibliography
\def\thebibliography{\DeclareRobustCommand{\VAN}[3]{##3}\VANthebibliography}

\usepackage{graphicx}	
\usepackage{amsmath}	
\usepackage{fancyref}
\usepackage{xcolor}
\usepackage{subcaption} 

\title[SWEEPS I. First detection]{Synoptic Wide-field EVN--\emph{e}-MERLIN Public Survey (SWEEPS) -- I. First steps towards commensal surveys with VLBI}

\author[C. Herb\'{e}-George et al.]{
C\'{e}lestin Herb\'{e}-George,$^{1,2}$\thanks{E-mail: herbegeorge@astro.rug.nl}
J. P. McKean,$^{1,2,3}$ 
Raffaella Morganti$^{1,4}$ 
and Jack F.~Radcliffe$^{2,5,6}$
\\
$^{1}$Kapteyn Astronomical Institute, University of Groningen, Landleven 12, 9747 AD Groningen, The Netherlands\\
$^{2}$Department of Physics, University of Pretoria, Lynnwood Road, Hatfield, Pretoria, 0083, South Africa\\
$^{3}$South African Radio Astronomy Observatory (SARAO), P.O. Box 443, Krugersdorp 1740, South Africa\\
$^{4}$ASTRON, Institute for Radio Astronomy, Oude Hoogeveensedijk 4, 7991 PD Dwingeloo, The Netherlands\\
$^{5}$Jodrell Bank Centre for Astrophysics, University of Manchester, Oxford Road, Manchester M13 9PL, United Kingdom\\
$^{6}$National Institute for Theoretical and Computational Sciences (NITheCS) South Africa \\
}

\date{Accepted 2024 December 02. Received 2024 November 28; in original form 2024 November 18}

\pubyear{2024}

\begin{document}
\label{firstpage}
\pagerange{\pageref{firstpage}--\pageref{lastpage}}
\maketitle

\begin{abstract}
The high angular resolution and sensitivity of Very Long Baseline Interferometry (VLBI) offer a unique tool to identify and study active galactic nuclei and star-formation activity over cosmic time. 
However, despite recent technical advances, such as multiple phase centre correlation, VLBI surveys have thus far been limited to either a few well-studied deep-fields or wide-areas to a relatively shallow depth. To enter the era of extensive statistical studies at high angular resolution, a significantly larger area of the sky must be observed to much better sensitivity with VLBI. The Synoptic Wide-field EVN--\emph{e}-MERLIN Public Survey (SWEEPS) is a proposed commensal observing mode for the EVN and \emph{e}-MERLIN, where single-target principle investigator-led observations are re-correlated at the position of known radio sources within 12 arcmin of the pointing centre. Here, we demonstrate a proof-of-concept of this methodology by detecting a 5.6\,mJy core-jet object at 1.7 GHz that would have otherwise been lost from the parent data set. This is the first object to be recovered as part of the SWEEPS pilot programme, which highlights the potential for increasing sample sizes of VLBI-detected radio sources with commensal observing modes in the near future.
\end{abstract}

\begin{keywords}
surveys -- techniques: high angular resolution -- techniques: interferometric -- galaxies: active -- radio continuum: galaxies
\end{keywords}

\section{Introduction}

Very Long Baseline Interferometry (VLBI) currently provides the sharpest view (3 to 200 mas at 1.7 GHz) of our Universe at radio wavelengths. Being sensitive to compact structure and unobscured by dust, the synchrotron emission from Active Galactic Nuclei (AGN) is readily detectable with sensitive $\umu$Jy-level VLBI observations \citep[e.g.][]{Kellermann_1998}. In addition, the radio emission from supernova remnants can be resolved and used as a radio-wavelength tracer for star formation in nearby galaxies \citep[e.g.][]{McDonald_2001}. The role of AGN feedback on star formation by either enhancement \citep[e.g.][]{Young_1981, Zhuang_2021} or suppression \citep[e.g.][]{Bower_2006, King_2015} can be revealed with VLBI \citep[][]{Middelberg_2013}.  This makes VLBI well-suited to study the star formation and assembly histories of galaxies, and can reveal the role of AGN feedback. Separately, the high angular resolution can confirm candidate dual or binary supermassive black holes, as predicted by hierarchical galaxy formation models, through direct observations of multiple compact cores \citep[e.g.][]{Rodriguez_2006}. Furthermore, high angular resolution imaging of gravitational lenses can test different dark matter models \citep[e.g.][]{Spingola_2018,Powell_2023}. 

Despite these high-impact science use cases, VLBI is still predominantly a single-object-focused instrument, with limited emphasis on large statistical studies.  Historically, this has been due to the prohibitive computational requirements of imaging the full ($\sim$7.5 arcmin in radius) field of view, albeit with some early successes \citep[e.g.][]{Garrett_2005}. In recent years, wide-field VLBI has become feasible with the advent of software correlators and multiple phase centre correlation capabilities \citep{Deller_2011, Keimpema_2015}. This technique has been successfully used to produce deep wide-field images of radio sources within a single or a few fields \citep[e.g.][]{Middelberg_2011,Herrera_Ruiz_2017,Radcliffe_2018,Njeri_2022}, but also shallow wide-field imaging over several hundred square degrees \citep{Deller_2014}. These surveys have started to bridge the gap between the theories developed with the single-object case studies and their applications to large populations of objects. However, correlating the full field of view with multiple phase centre correlation remains the exception. 

Leveraging multiple phase centre correlation in the form of dedicated projects, such as the mJy Imaging Very Long Baseline Array Exploration survey (mJIVE--20; \citealt{Deller_2014}), requires large amounts of observing time, which is a limited resource with VLBI instruments [e.g., the European VLBI Network (EVN), including the enhanced Multi-Element Remotely Linked Interferometer Network (\emph{e}-MERLIN), observes about 63 days of the year, of which around 30 days are at 1.7 GHz]. Fortunately, when combined with multiple phase centre correlation, interferometry perfectly lends itself to a commensal survey mode, which we aim to establish through the Synoptic Wide-field EVN--\emph{e}-MERLIN Public Survey (SWEEPS; PI: McKean). In this project, the data from standard single-target principle investigator (PI)-led observations can be re-correlated to the position of known radio sources within the primary beam of the telescopes, with no impact on the original PI-led experiment. Initially, the positions that need to be known apriori can be sourced from large radio surveys, such as the LOFAR Two Metre Sky Survey (LoTSS; \citealt{Shimwell_2022}), the Faint Images of the Radio Sky at Twenty-one centimetres (FIRST; \citealt{Becker_1995}) or the Very Large Array Sky Survey (VLASS; \citealt{Lacy_2020}). Alternatively, VLBI networks with a compact cluster of telescopes, such as \emph{e}-MERLIN, can also produce a low-resolution, but wide-field image, from which the source positions can be found. This removes the ancillary data requirements and can increase processing efficiency by only re-correlating at the position of those objects visible at the particular frequency of the VLBI observation. Alternatively, if there are sufficient compute resources, the primary beam of the array could be fully gridded with multiple phase centres, irrespective as to whether there is a known radio source or not.

\par In this letter, we demonstrate the proof-of-concept for such a commensal survey mode by using the calibrator field of a standard PI-led VLBI observation with the EVN and \emph{e}-MERLIN. A bright source close to the pointing centre is selected to investigate its morphology and properties, showcasing the scientific analysis that can be done in the future with SWEEPS. Throughout, we assume a flat $\Lambda$CDM cosmology, with H$_0 = (67.4 \pm0.5)\,$km$\,$s$^{-1}\,$Mpc$^{-1}$, $\Omega_m = 0.315 \pm0.007$ and $\Omega_\Lambda = 0.6847 \pm 0.0073$ \citep{Planck_2020}. We also assume that the spectral index, $\alpha$, is given by $S_\nu \propto \nu^\alpha$, where $S_\nu$ is the flux density and $\nu$ is the frequency.

\section{Observations and data reduction}
\subsection{Observing setup}

 The parent EVN and \emph{e}-MERLIN 18~cm/1.7~GHz observations used for this pilot project are of the lensed quasar SDSS\,J1004+4112 (target) and the phase calibrator J0948+4039 (Project EM149; PI: McKean). The target and phase calibrator were observed for a total of 24~h, split equally over two days, with nodding cycles of 3.5 and 1.5~min, respectively. Additionally, two bright fringe-finders, 4C39.25 and DA\,193, were observed for a total of 33~min, split over 7 scans. The data were recorded at 1~Gb\,s$^{-1}$ in $4\times 32$ MHz spectral windows with 128 channels each (equivalent to a channel resolution of $250$~kHz), centred at 1658.49~MHz. Currently, the \emph{e}-MERLIN stations record data at 512~Mb\,s$^{-1}$, meaning that only the central two spectral windows (with 128 channels each of width $250$~kHz) were available. 
 
 \subsection{Multiple phase centre correlation}
 
\begin{figure}
  \centering
  \includegraphics[width = 1\linewidth]{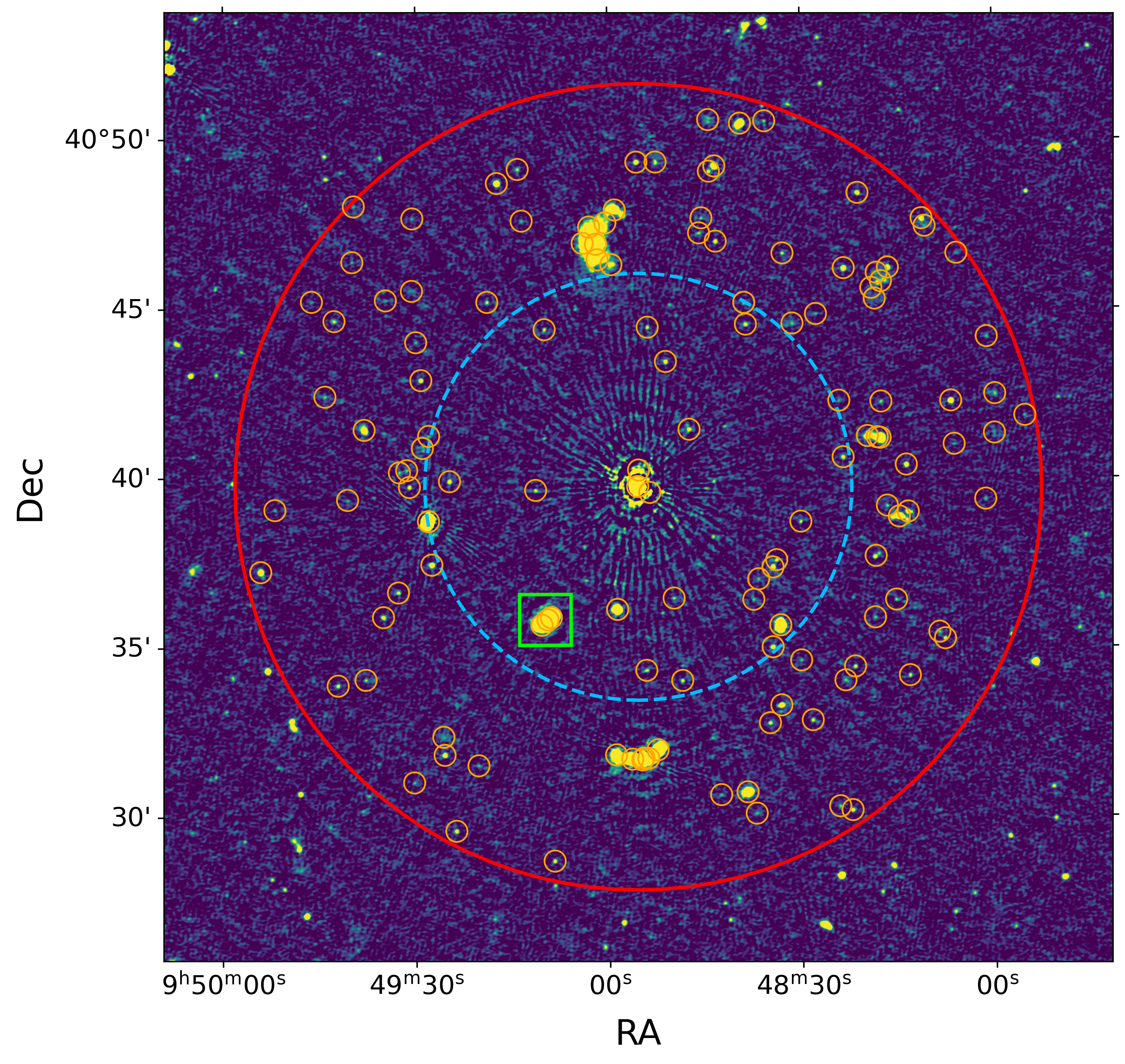}
  \caption{144~MHz LoTSS DR2 image of the J0948+4039 phase calibrator field. The orange circles of radius 18.8 arcsec show each phase centre and represent the average 10 per cent time and bandwidth smearing radius. The circles in dashed-blue (6.3 arcmin in radius) and in solid-red (11.9~arcmin in radius) correspond to the half-power beam widths of a 100 and 32~m telescope, respectively. The green box corresponds to SWEEPS J094909+403548, our first target from the project, which includes three phase centres across the LoTSS detection.}
  \label{fig:J0948_overview}
\end{figure}

 The raw baseband data from EM149 were kept on disk and re-correlated using the multiple phase centre mode of SFXC \citep{Keimpema_2015} with an internal channel width of 1.95~kHz (16\,384 channels per spectral window) and an integration time of $1.3056 \times 10^{-4}$~s. These parameters ensure that bandwidth and time smearing is kept to 1 per cent at the half-power beam width of a 32~m dish, which is 11.9 arcmin in radius from the delay centre. Since the phase centres need to be known before the re-correlation, the LoTSS data release 2 (DR2; \citealt{Shimwell_2022}) was used to provide the coordinates of the sources within the field. Although at a lower frequency of 144~MHz, LoTSS provides the deepest (median rms sensitivity of $83\,\umu$Jy\,beam$^{-1}$) wide-area radio data available for our study.  
 We selected 130 phase centres in the calibrator field, which are shown in Fig.~\ref{fig:J0948_overview}. For some select extended sources, additional phase centres were manually added to test the imaging capabilities of the shorter baselines of \emph{e}-MERLIN within the combined EVN and \emph{e}-MERLIN array. Each phase centre was outputted with a channel width of 250~kHz and 2~s integrations, corresponding to 10 per cent smearing radii of 20.4 and 17.2 arcsec for bandwidth and time, respectively. Unfortunately, due to a bug in the SFXC code (which has since been corrected),  only one-quarter of the integrations were accumulated during the multiple phase centre correlation (Campbell, priv. comm.), resulting in a doubling of the rms noise with respect to the theoretical expectation for the data presented here. The EVN project code for the resulting multiple phase centre correlation dataset is EM160 (PI: McKean).

\subsection{Data reduction and calibration}

The output dataset was retrieved from the EVN archive\footnote{http://archive.jive.nl/} and reduced using the Common Astronomy Software Applications (CASA\footnote{https://casa.nrao.edu/}; \citealt{CASA_2022, van_Bemmel_2022}). The visibilities of the phase calibrator and fringe-finders were inspected for any erroneous data and radio frequency interference (RFI), which were flagged where appropriate. The antenna gain curves and measured system temperatures were used to scale the amplitudes to a physically meaningful flux density.  The instrumental delays were corrected by selecting a 1~min interval during a scan of 4C39.25, and the bandpass correction was calculated using all four scans of 4C39.25. The multi-band delays were corrected by computing the delays and delay rates every 60~s between Effelsberg and all other antennas for the phase calibrator. After applying these calibration tables, three rounds of phase-only self-calibration with solution intervals of 120, 60 and 30~s were conducted, followed by three rounds of amplitude and phase self-calibration with the same solution intervals. The resulting image (Briggs weighting; Robust = 0) of the phase calibrator had a peak surface brightness of 565~mJy\,beam$^{-1}$ and an off-source rms of 252~$\umu$Jy\,beam$^{-1}$. The nearest bright source from the pointing centre that was detected by LOFAR and FIRST was selected as our test object, which is shown by the cyan box in Fig.~\ref{fig:J0948_overview}. Finally, the calibration tables and data flags were applied to our target dataset, which included three phase centres across the source, and the primary beam correction was computed and applied using a modified version of the {\sc vpipe} pipeline \citep{Radcliffe_vpipe_2024}. The three phase centres were then imaged using a Briggs (Robust = 0) and natural weighting of the visibilities.

\section{Results and analysis}

\subsection{First VLBI detection from a commensal observation}
\label{sec:VLBI_detection}
\begin{figure}
 \centering
  \includegraphics[width = \linewidth]{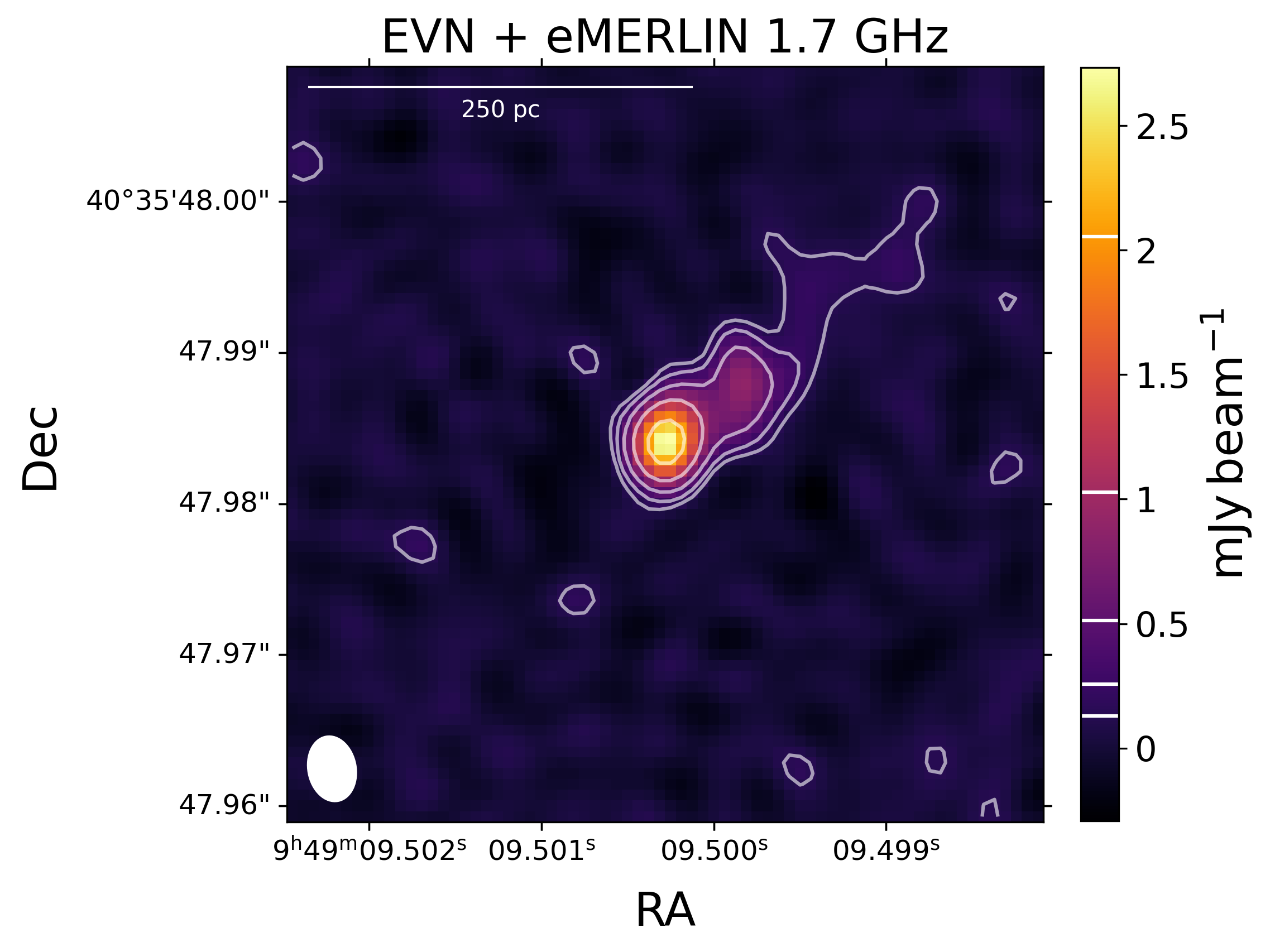}
\caption{Briggs weighted (Robust = 0) 1.7~GHz EVN and \emph{e}-MERLIN image of SWEEPS~J094909+403548, with a restoring beam of 4.3~mas $\times$ 3.1~mas at a position angle of 10.7~deg east of north. The peak surface brightness is 2.73~mJy~beam$^{-1}$ and the off-source rms is 64.9 $\umu$Jy~beam$^{-1}$. The contour levels are at (2, 4, 8, 16, 32)~$\times~\sigma_{\text{rms}}$, where $\sigma_{\text{rms}}$ is the rms map noise.}
\label{fig:EVN_SWEEPS01}
\end{figure}

The Briggs-weighted image of the central phase centre of our target dataset is shown in Fig. \ref{fig:EVN_SWEEPS01}, which reveals SWEEPS~J094909+403548. This detection provides a proof-of-concept for the SWEEPS project and for a commensal observing mode with VLBI. The object has a core-jet like structure, with a compact core that has a naturally weighted peak surface brightness of 3.83\,mJy\,beam$^{-1}$, and another compact component and faint diffuse emission extending to the north-west. From fitting two elliptical Gaussian model components in the image-plane, we find that the object has a total flux density of 5.60~mJy in the naturally-weighted image. The results of the fit are shown in Table \ref{tab:VLBI_source_properties}.
\begin{table*}
    \centering
    \caption{EVN and \emph{e}-MERLIN Gaussian model fitting results. $\theta_\mathrm{maj}$ and $\theta_\mathrm{min}$ are the convolved Gaussian model major and minor axes, respectively.}
    \label{tab:VLBI_source_properties}

    \begin{tabular}{lccccccr}
        \hline
         {Weighting} & {Component} & {RA} & {Dec} & {$\theta_\mathrm{maj}$} & {$\theta_\mathrm{min}$} & {Flux density} & {Peak surface brightness} \\
         & & {(deg)} & {(deg)} & {(mas)} & {(mas)} & {(mJy)} &{(mJy\,beam$^{-1}$)} \\
        \hline
Natural & Core & 147.289584 & +40.596663 & 10.2 & 8.7 &  3.83 $\pm$ 0.12 &  3.49 $\pm$ 0.06\\
Natural & NW & 147.289581 & +40.596664 & 19.5 & 9.2 &  1.77 $\pm$ 0.19 &  0.79 $\pm$ 0.06\\
Briggs & Core & 147.289584 & +40.596662 & 4.2 & 3.3 &  2.84 $\pm$ 0.12 &  2.70 $\pm$ 0.07\\
Briggs & NW & 147.289583 & +40.596663 & 7.5 & 4.5 &  2.14 $\pm$ 0.24 &  0.85 $\pm$ 0.07\\

        \hline
    \end{tabular}
\end{table*}

Interestingly, this object was observed as part of the mJIVE--20 survey \citep{Deller_2014}, but no core was detected despite the object having a peak surface brightness of at least 25 times the detection threshold of that survey. The non-detection was likely due to the search region, a $3.5\times3.5$ arcsec box centred on the brightest pixel of the FIRST image, being too small. That position is 7.1 arcsec away from the peak surface brightness of our detection. This presents an important issue for VLBI surveys, where the extended emission detected from lower-resolution imaging may misguide the phase centre selection. This can be mitigated using optical ancillary data or by making sufficiently wide-field images from the shorter-baselines of the interferometric array, as is the plan for SWEEPS.

\subsection{Ancillary radio data and spectral index map}

\begin{figure*}
 \centering
  \includegraphics[width = 1\linewidth]{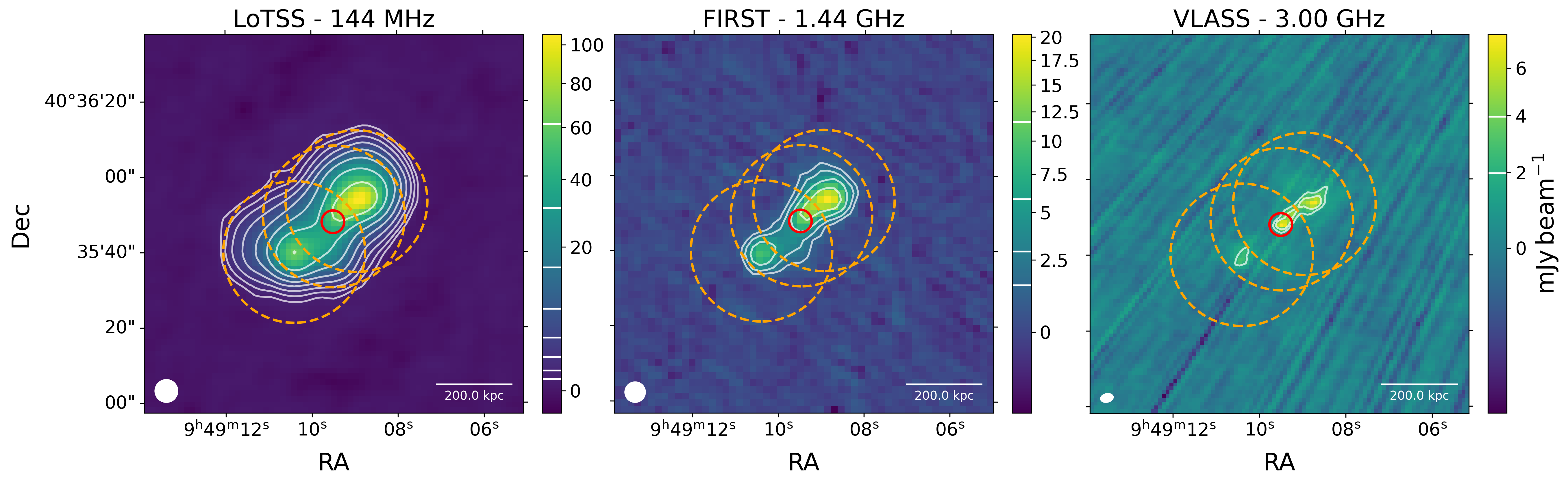}
\caption{The ancillary radio images of SWEEPS J094909+403548. From left to right: LoTSS DR2 144~MHz image with a beam size of $6\times6$ arcsec, the FIRST 1.44 GHz image with a beam size of $5.4\times5.4$~arcsec and the VLASS epoch two 3~GHz image with a beam size of $3.3\times2.2$~arcsec. Overlaid are the surface brightness contours at $(5, 10, 20, 40, 80, 160, 320, 640)\times\sigma_{\text{rms}}$, where $\sigma_{\text{rms}}$ is the rms map noise of each image. The centre of the red circle indicates the location of the SWEEPS J094909+403548 VLBI detection, shown in Fig. \ref{fig:EVN_SWEEPS01}, and the orange dashed circles show the 18.8~arcsec 10 per cent average smearing radius of the three phase centres used for this object.}
\label{fig:ancillary_images}
\end{figure*}

We now investigate the properties of SWEEPS J094909+403548 using the available ancillary data. The radio surveys covering this field are the LOFAR LoTSS DR2 (144 MHz), the FIRST survey (1.44 GHz) and VLASS epochs one and two (3 GHz). VLASS is still ongoing, meaning the final combined data products are not yet available. The VLASS epoch two image was selected as this had the best noise properties. The images from these surveys for SWEEPS J094909+403548 are shown in Fig. \ref{fig:ancillary_images}.

We see that the VLBI detection is associated with a large-scale radio source that is extended by around 50 arcsec in the north-west to south-east direction on the sky, which is suggestive of two lobes. Interestingly, at these angular resolutions and at the lower frequencies, the VLBI position is not coincident with the peaks in surface brightness at the edge of the radio emission, but is instead located in between them. Also, the small-scale extension seen on VLBI-scales is in the direction of the north-western brightness peak. We also note that the south-east component shows evidence of emission towards the west at 144~MHz and at 1.44 GHz. It is not clear if this is due to back flow from the AGN plasma, or perhaps evidence for another radio source (see below).

\begin{figure}
 \centering
  \includegraphics[width = \linewidth]{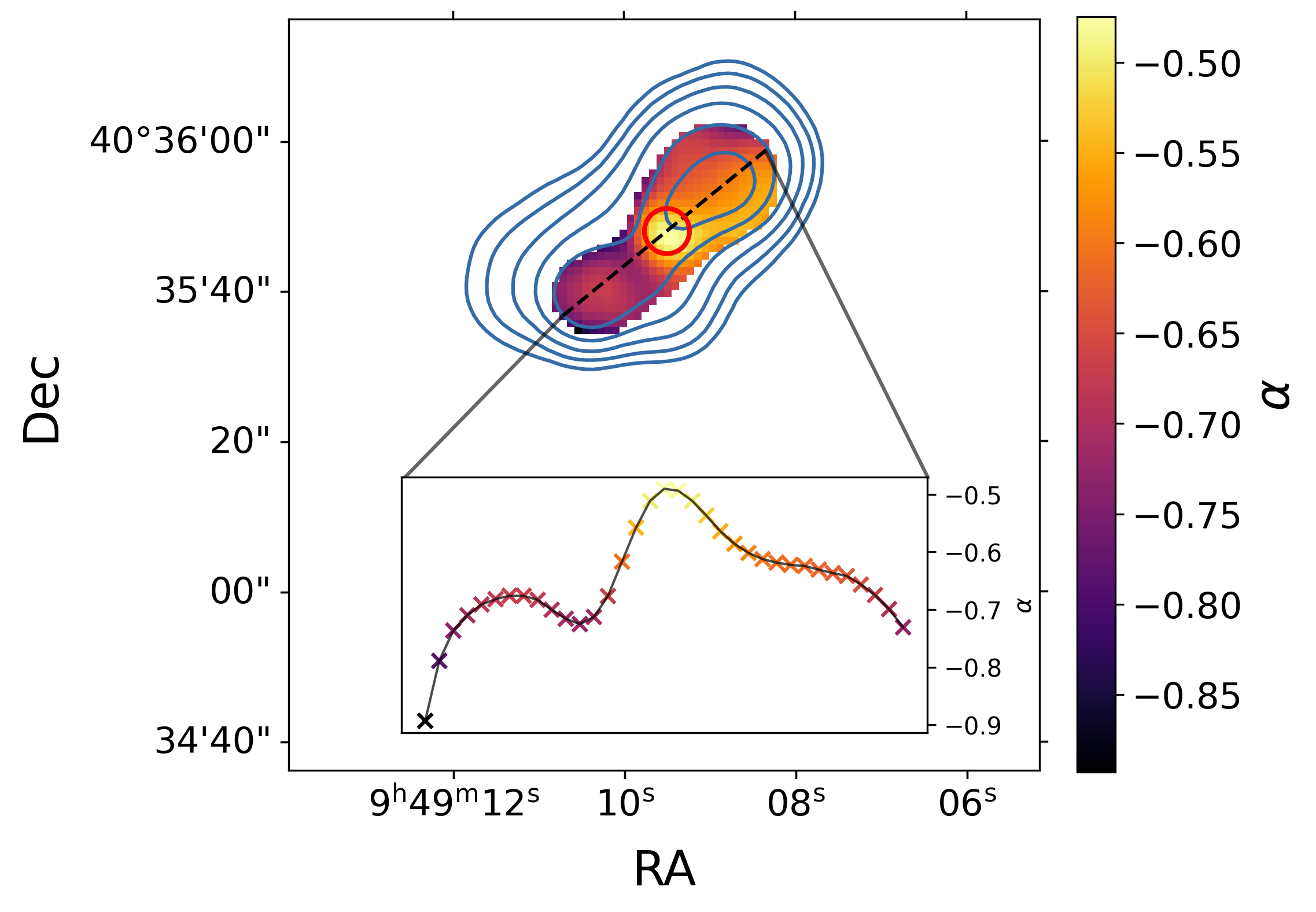}
\caption{The spectral index map from 144~MHz to 3~GHz, using the available ancillary radio data presented in Fig. \ref{fig:ancillary_images}. Here, the FIRST and VLASS data were smoothed to the same beam size as the LOFAR imaging ($6\times6$~arcsec). The contours represent the LOFAR surface brightness at (20, 40, 80, 160, 320, 640)$\times\sigma_{\rm rms}$, where $\sigma_{\rm rms}$ is the rms map noise. The black dotted line shows a transect of the spectral index across the emission, with the line profile of the spectral index shown. The centre of the red circle indicates the location of the SWEEPS J094909+403548 VLBI detection.}
\label{fig:spec_idx_map}
\end{figure}

In order to investigate the spectral properties of the overall structure, we smoothed the images to a common beam size of $6\times6$~arcsec and made a spectral index map, which is shown in Fig.~\ref{fig:spec_idx_map}. We can see there is a flat spectrum region that is coincident with the EVN and \emph{e}-MERLIN detection at 1.7 GHz, with a spectral index of $\alpha_{\rm core} = -0.48$. Therefore, the location and spectral index of the VLBI detection is consistent with what is expected for an AGN core. The two regions towards the north-west and south-east exhibit a steepening of the spectral index, with $\alpha_{\rm nw} = -0.61$ and $\alpha_{\rm se} = -0.64$, respectively. This is typical of FR I radio galaxies \citep{de_Gasperin_2018}. We also investigated the presence of a break in the spectral index. However, due to the quality and the different resolutions of the data, the results were too uncertain to ascertain the presence of a spectral break, and so, the age of the plasma is unknown.

To measure the flux density of the various parts of the radio emission, we fitted Gaussian model components in the image-plane at each frequency. Three Gaussian model components were used, two components for the north-western extension and a third for the south-eastern extension. The resulting fit parameters are shown in Table \ref{tab:ancillary_radio_properties} in Appendix \ref{sec:Fit parameters}.

\subsection{Optical and infrared data}

\begin{figure}
 \centering
  \includegraphics[width =\linewidth]{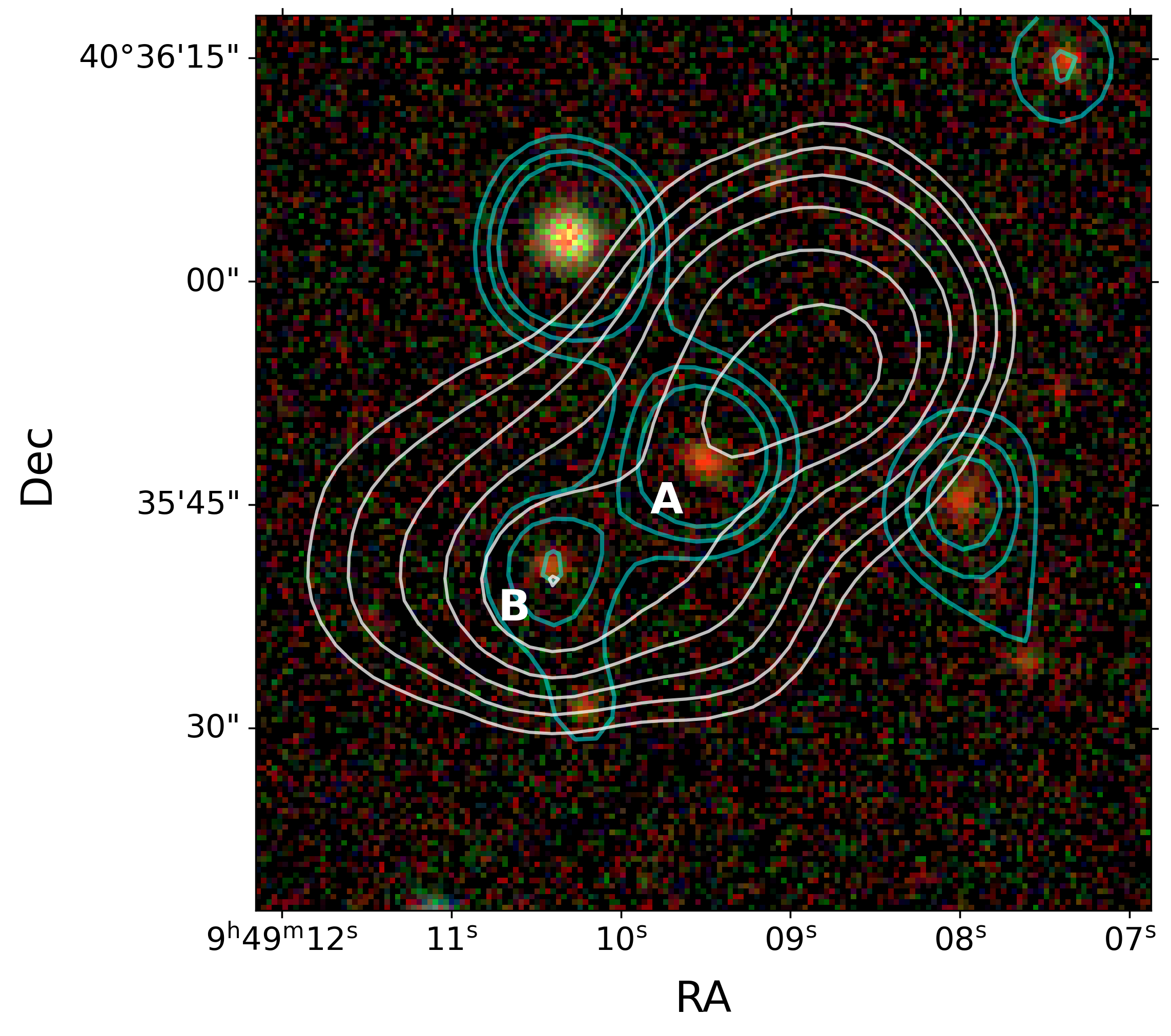}
\caption{A pseudo-colour image of the SWEEPS J094909+403548 field using the $g$-, $r$- and $i$-band SDSS images. The LOFAR contours from Fig. \ref{fig:ancillary_images} are overlaid in white. The WISE 3.4~$\umu$m contours are overlaid in cyan, revealing the infrared emission from the different sources. The likely host galaxy of SWEEPS J094909+403548 is labelled as A, and a second optical galaxy that coincides with the south-eastern radio component is labelled as B.}
\label{fig:SDSS_RGB}
\end{figure}

To further understand the nature of SWEEPS J094909+403548, we have created a pseudo-colour optical image using the $g$-, $r$- and $i$-band images of the field from the Sloan Digital Sky Survey (SDSS; \citealt{SDSS_DR17}), which is shown in Fig. \ref{fig:SDSS_RGB}. In addition, we also over-plot the contours from the Wide-Infrared Survey Explorer (WISE; \citealt{Wright_2010}) W1 at 3.4~$\umu$m. 

We find two optical detections that coincide with the radio emission seen by the radio surveys. One optical object, SDSS J094909.51+403548.0, which we denote as object A, is at the location of SWEEPS J094909+403548, suggesting it is the host galaxy of the AGN core detected at VLBI scales. There is no optical emission coincident with the north-western radio component. Object A was also observed during the BOSS spectrographic survey \citep{BOSS_survey_2013}, finding a redshift of $z = 0.52426\pm0.00013$. The second object, SDSS J094909.51+403548.0, which we denote as object B, is coincident with the south-eastern radio extension. There is no recorded spectroscopic redshift for this object, but there is an SDSS photometric redshift of $z=0.514\pm0.038$. We also see that there are W1 detections at the location of objects A and B.

\subsection{Properties of SWEEPS J094909+403548} \label{sec: properties of SWEEPS J094909+403548}
With the redshift and spectral index of object A known, the monochromatic 1.7~GHz spectral power of the VLBI component can be found using, 
\begin{equation}
    L_{1.7~\text{GHz}} = \frac{4\pi d^2_\text{L}}{(1+z)^{1+\alpha}} S_\text{obs}\,, \label{eq:radio_power}
\end{equation}
where $d_\text{L}$ is the luminosity distance, $z$ is the redshift, $\alpha$ is the spectral index, and $S_\text{obs}$ is the observed flux density. Using the naturally-weighted values presented in Table \ref{tab:VLBI_source_properties}, the monochromatic radio power for SWEEPS J094909+403548 is found to be $(5.13\pm 0.55)\times10^{24}$~W\,Hz$^{-1}$.

Similarly, the brightness temperature, $T_\text{b}$, of the object can be calculated using,
\begin{equation}
    T_\text{b} = 1.22 \times 10^{12}(1+z) \left(\frac{S_\text{obs}}{1\, \text{Jy}}\right) \left(\frac{\nu}{1\,\text{GHz}}\right)^{-2}\left(\frac{\theta_{\text{maj}}\theta_\text{min}}{1\,\text{mas}^2}\right)^{-1},
    \label{eq:brightness_temp}
\end{equation}
where $\theta_{\text{maj}}$ and $\theta_{\text{min}}$ are the major and minor axis, respectively, of the fitted Gaussian to the compact core \citep{condon_1982, Radcliffe_2018}. Here, the core of the Briggs-weighted image was used to fit the Gaussian, as this weighting scheme is less sensitive to extended emission, which could be associated with the jet. A measured flux of 2.84 mJy was found to be enclosed within a 4.23$\times$3.33 mas region centred on the brightest pixel of Fig. \ref{fig:EVN_SWEEPS01}. The resulting brightness temperature of SWEEPS J094909+403548 is found to be $(1.36 \pm 0.15)\times10^8$~K. As discussed by \citet{Condon_1991}, radio emission from AGN activity typically has a brightness temperature of $>10^5$ K. This, in combination with the core-jet morphology, suggests that this object must be an AGN. This relatively simple analysis highlights the usefulness of VLBI in the context of interpreting the emission from low-resolution all-sky surveys.

\subsection{A second radio galaxy?}

The optical and infrared emission from object B (see Fig. \ref{fig:SDSS_RGB}) is coincident with the radio emission from the south-eastern component. However, no VLBI detection is present in the region of object B. The image rms at the location of the optical detection is $31~\umu$Jy~beam$^{-1}$, which, if we adopt a detection threshold of 6$\sigma$, gives an upper limit of $186~\umu$Jy~beam$^{-1}$ for the surface brightness of any VLBI counterpart to the low resolution radio emission. We note that within the phase centre, the peak brightness is 167 $\umu$Jy\,beam$^{-1}$. The steep spectral index of the south-eastern region would also be consistent with a non-detection on VLBI-scales as it indicates optically thin emission, which is generally extended. Using equation~(\ref{eq:radio_power}) and the flux density measured at 1.44 GHz from FIRST and the photometric redshift of $z=0.514\pm0.0378$ from SDSS, the monochromatic radio power of the south-eastern component is found to be ($1.65\pm 0.19)\times10^{25}$ W\,Hz$^{-1}$. This is certainly consistent with AGN emission. We also note that as part of the mJIVE--20 survey, around 20 per cent of FIRST sources had VLBI detections. Therefore, although we find no evidence of a radio core associated with object B, whether or not SWEEPS J094909+403548 is the core of a double-sided radio source, or a core-jet source with a companion radio galaxy to the south-east is not clear at this stage.

We also searched the full field-of-view of the three phase centres shown in Fig.~\ref{fig:ancillary_images}. However, no further VLBI-detections above a local rms of $6\sigma$ were found.

\section{Conclusions}

In this letter, we have shown a proof of concept for a commensal wide-field VLBI survey using the multiple phase centre technique available in modern software correlators. Using a PI-led observation at 1.7~GHz with the EVN and \emph{e}-MERLIN VLBI network that had a different science goal, we re-correlated the data using 130 phase centres selected from the 144~MHz LoTSS DR2 parent catalogue. We selected a bright object near the pointing centre of the observation, reduced the data and produced robustly and naturally weighted images. A source with a naturally weighted peak surface brightness of 3.75\,mJy\,beam$^{-1}$ and a total flux density of 5.61~mJy was detected. This object has bright and compact components with extended emission towards the north-west that is morphologically similar to an AGN core and a jet. This is supported by the ancillary data, which shows a large lobe to the north-west at lower angular resolution. No VLBI detection is found at the location of the south-eastern lobe seen in the low-resolution ancillary data. However, a galaxy is detected in the optical and infrared ancillary data at this location.

Crucially, the detection presented here was obtained without additional observing time or correlator capabilities, and used standard data reduction routines. With this proof-of-concept in hand, the other 260 phase centres from our pilot project will be processed and analysed in an upcoming paper. There, we will discuss the expected number of sources that SWEEPS will detect, and the specific routines that need to be developed, such as baseline-dependent source subtraction and the calibration of the shorter \emph{e}-MERLIN baselines.

\section*{Acknowledgements}
We would like to thank the anonymous referee for their valuable comments. We would also like to thank Bob Campbell, Marjolein Verkouter and Suma Murthy for their help with the re-correlation of the data and the helpful discussions. CHG and JPM acknowledge support from the Netherlands Organization for Scientific Research (NWO) (Project No. OCENW.XS23.1.133). This work is based on the research supported in part by the National Research Foundation of South Africa (Grant Number: 128943). The European VLBI Network is a joint facility of independent European, African, Asian, and North American radio astronomy institutes. Scientific results from data presented in this publication are derived from the following EVN project code(s): EM160. \emph{e}-MERLIN is a National Facility operated by the University of Manchester at Jodrell Bank Observatory on behalf of STFC.

\section*{Data Availability}
The data used in this project are publicly available via the EVN archive hosted at JIV-ERIC and are associated with project EM160 (PI: McKean).

\bibliographystyle{mnras}
\bibliography{references}

\appendix

\section{Ancillary Radio data fit parameters}\label{sec:Fit parameters}

\begin{table}

\centering
\caption{Gaussian model fitting results for each radio survey. $\theta_\mathrm{maj}$ and $\theta_\mathrm{min}$ are the convolved Gaussian model major and minor axes, respectively.}
\label{tab:ancillary_radio_properties}
\begin{subtable}{0.45\textwidth}
    \centering
    \caption{LoTSS}
    
    \begin{tabular}{lccccr}
        \hline
        Component & {RA} & {Dec} & {$\theta_\mathrm{maj}$} & {$\theta_\mathrm{min}$} & {Flux density} \\
         & {(deg)} & {(deg)} & {(arcsec)} & {(arcsec)} & {(mJy)} \\
        \hline
NWa & 147.2885 & +40.5979 & 20.4 & 12.1 &  364 $\pm$ 9 \\
NWb & 147.2866 & +40.5983 & 11.1 & 7.2 &  142 $\pm$ 4 \\
SE & 147.2932 & +40.5945 & 12.1 & 10.0 &  179 $\pm$ 6 \\

        \hline
    \end{tabular}
\end{subtable}

\begin{subtable}{0.45\textwidth}
    \centering
    \caption{FIRST}
    \label{tab:FIRST_properties}
    \begin{tabular}{lccccr}
        \hline
        Component & RA & Dec & $\theta_\mathrm{maj}$ & $\theta_\mathrm{min}$ & Flux density \\
        & (deg) & (deg) & (arcsec) & (arcsec) & (mJy) \\
        \hline
NWa & 147.2886 & +40.5976 & 14.9 & 6.6 &  44 $\pm$ 2 \\
NWb & 147.2864 & +40.5982 & 7.0 & 5.7 &  21 $\pm$ 1 \\
SE & 147.2932 & +40.5943 & 7.6 & 7.0 &  17 $\pm$ 1 \\
        \hline
    \end{tabular}
\end{subtable}

\begin{subtable}{0.45\textwidth}
    \centering
    \caption{VLASS}
    \label{tab:VLASS_properties}
    \begin{tabular}{lccccr}
        \hline
        Component & RA & Dec & $\theta_{maj}$ & $\theta_{min}$ & Flux density \\
         & (deg) & (deg) & (arcsec) & (arcsec) & (mJy) \\
        \hline
NWa & 147.2893 & +40.5968 & 4.8 & 3.0 &  13 $\pm$ 1 \\
NWb & 147.2867 & +40.5983 & 9.0 & 4.7 &  31 $\pm$ 3 \\
SE & 147.2929 & +40.5945 & 11.1 & 6.1 &  18 $\pm$ 4 \\
        \hline
    \end{tabular}
\end{subtable}
\end{table}

\bsp	
\label{lastpage}
\end{document}